\documentstyle[12pt,psfig,epsf]{article}
\newcommand{\ezero}{\setcounter{equation}{0}}

\newcommand {\oa} {\mbox{${\cal O}(\alpha)$}}
\newcommand{\nn}{\noindent}

\newcommand{\bq}{\begin{equation}}
\newcommand{\eq}{\end{equation}}
\newcommand{\ba}{\begin{eqnarray}}
\newcommand{\ea}{\end{eqnarray}}
\begin{document}
\begin{titlepage}
\bigskip
\noindent
IC/94/203
 
\noindent
TAUP-2183-94
 
\noindent
 August  1994
 
\vspace*{.50cm}
{ \huge 
\bf \begin{center}
 
  Radiative Corrections for
  Pion Polarizability  Experiments
 
 \vspace*{.4cm}
\end{center}
   }
\vspace*{.8cm}
\begin{center}
{\large \bf A.~A.~Akhundov$^{1,2}$, S.~Gerzon$^{3}$,
 S.~Kananov$^{3}$,  M.~A.~Moinester$^{3}$\\}
 
\end{center}
\vspace*{.7cm}
\begin{center} {\it
$\: ^{1}$ International Centre for Theoretical Physics, 34014 Trieste, Italy \\
 $\: ^2$ Institute of  Physics, Azerbaijan Academy of Sciences,
         370143 Baku, Azerbaijan \\
 $\:^3$ Raymond and Beverly Sackler Faculty of Exact Sciences, \\
        School of Physics and Astronomy, Tel Aviv University, \\
        69978 Ramat Aviv, Israel  \\
   } \end{center}

\vspace*{.7cm}
\vfill
\thispagestyle{empty}
\normalsize
\centerline{\bf Abstract}
\vspace*{.2cm}
 
\small
\nn
 
 We use the semi-analytical program RCFORGV
to evaluate radiative corrections to
one-photon radiative  emission 
in the high-energy scattering of pions 
in the Coulomb field of a nucleus with atomic number Z.
It is shown that radiative corrections can simulate
a pion polarizability effect. The average 
effect is $\alpha_{\pi}^{rc}=-\beta_{\pi}^{rc}
=(0.20 \pm0.05)\times10^{-43}cm^3$,
for pion energies 40-600 GeV. We also 
study the range of applicability of the 
equivalent photon approximation in describing 
one-photon radiative emission. 
 
\normalsize
 
\vfill
\end{titlepage}
\newpage
\newpage
\section
{Introduction \label{intr}}
\ezero
 
Pion electric $\bar{\alpha}$  and magnetic $\bar{\beta}$
Compton polarizabilities~\cite{kle}-\cite{lvov3} characterize
 the induced transient dipole
moments of pion subjected to external oscillating electric $\vec{E}$ and
magnetic $\vec{H}$ fields. The polarizabilities can be
obtained from precise measurements of the gamma-pion Compton scattering
differential cross section. They probe the rigidity of the internal structure
of the pion, since they are induced by the rearrangement of the
pion constituents via action of the  photon electromagnetic fields during
scattering.
For the $\gamma\pi$ interaction at low energy, chiral symmetry
provides a rigorous way to make predictions. This approach~\cite{hols1,dono}
 yields $\bar\alpha_{\pi}$ = -$\bar\beta_{\pi}$ = 2.7
$\pm$ 0.4 (in units $10^{-43}~cm^3$).

The radiative scattering of high energy pions in the
Coulomb field of a nucleus~\cite{GMOP}:
$$
\pi(p_1) + Z(p) \rightarrow \pi(p_2) + Z(p') + \gamma(k'),
\eqno(1)
$$
determines the  $\gamma\pi$ Compton scattering:
$$
\pi(p_1) + \gamma(k) \rightarrow \pi(p_2)  + \gamma(k').
\eqno(2)
$$
The $\gamma\pi$ scattering was measured~\cite{Antip12,Antip34} with 40
 GeV pions at Serpukhov via the Primakoff radiative
scattering process (1). The
incident pion scatters from a virtual photon, characterized by
4-momentum $k=p-p'$; and the final state gamma ray
and pion are detected in coincidence.
The pion electric polarizability $\bar{\alpha_{\pi}}$ was deduced
in a low statistics ($\sim$ 7 000 events) experiment to be:
$$
\bar\alpha_{\pi} = -\bar\beta_{\pi} = 6.8 \pm
{1.4}_{stat} \pm {1.2}_{syst}.
\eqno(3)
$$
It was assumed in the analysis that ${{\bar{\alpha}_\pi}} +
{{\bar{\beta}_\pi}} = 0$, as expected theoretically~\cite{hols1}.
This result differs from the chiral prediction by more than two standard
deviations.
The experimental situation points to the need for much higher quality
data and more attention to the systematic uncertainties arising from
different measurement and analysis techniques. A new Primakoff experiment
is planned at FNAL~\cite{mur}.

\begin{figure}[htb]
\centerline{\psfig{figure=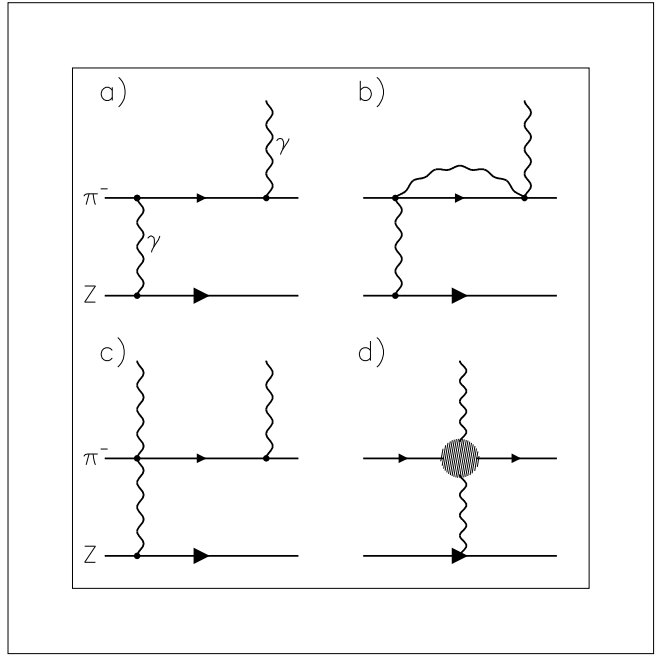,height=10cm}}
\caption{{\it Feynman graphs for radiative scattering of pions on
Coulomb center: a) the Born approximation;
b) the one-loop QED corrections;
c) the double photon bremsstrahlung;
) the structure effects - pion polarizabilities.}}
\label{fig1}
\end{figure}

In the lowest order, the process of the radiative scattering
 of the pions in the Coulomb field of a nucleus is
described by Feynman graphs of Fig.~\ref{fig1}:
\footnote{In Fig.~\ref{fig1}, only one of the full set of needed
diagrams is presented.}
 1a) the QED interaction
of the pointlike pion with the electromagnetic field of the nucleus;
1d) the two-photon interaction of the pion, including the structure effects
    described by the polarizabilities of the pion.
In the next order of $\alpha$,  there are contributions
from the QED processes  presented in Fig.~\ref{fig1}(b,c).
 The double bremsstrahlung process (1c):
$$
\pi(p_1) + Z(p) \rightarrow \pi(p_2) + Z(p') + \gamma(k_1)+\gamma(k_2),
\eqno(4)
$$
and the one-loop diagrams (1b) with the virtual photon, both
contribute to the measured cross section of the reaction (1).
 
The total cross section for the radiative scattering process (1)
can be written in the form:
 
$$
d {\sigma}^{tot} = d {\sigma}^{Born} + d {\sigma}^{pol} +
                         d {\sigma}^{rc}.
\eqno(5)
$$
Here, $d {\sigma}^{Born}$ corresponds to scattering of a pointlike pion
in the Born approximation, and the following terms give the contributions
of the structure effects (linear terms in the pion polarizability) and the
radiative corrections.
The radiative corrections are important for high statistics experiments,
and are needed to extract the Compton cross sections from data.
The experimental data can be interpreted in terms of polarizabilities,
only after accounting for the
radiative corrections.
 
The theoretical investigation~\cite{akhu1}-\cite{akhu3} of these radiative
corrections (RC) was
undertaken initially for the analysis of the experiment~\cite{Antip12,Antip34}
of JINR (Dubna) and IHEP (Serpukhov).
It was shown that the detailed properties of the gamma detector are
important; such as the
photon detector threshold, energy and position resolution, and
the two-photon angular resolution. For
the setup of Antipov {\em et al.}~\cite{Antip12,Antip34} at 40 GeV,
the radiative corrections  were estimated to
affect polarizability determinations at the level of $\sim 3\%$~\cite{akhu3}.

In this note we present a study of the radiative corrections to the
process (1) at higher energies and for higher Z targets. Our
calculations were carried out with the semi-analytical
program  RCFORGV~\cite{akhu4}.
 
\section
{The semi-analytical program RCFORGV
\label{RCFORGV}}
 
The analytical formulae for the RC to the inclusive spectra
of the scattering
of a spinless pointlike particle by a Coulomb center with emission of a hard
photon were obtained in~\cite{akhu1,akhu3}.
  The contribution of the QED one-loop corrections (Fig.~\ref{fig1}b) and
  of the double
 bremsstrahlung from pions (Fig.~\ref{fig1}c)
 were calculated exactly without any
 approximation and for $m_{\pi} \neq 0 $.
 The calculation of the Feynman diagrams was carried out using
the program SCHOONSCHIP~\cite{SCHOONSCHIP},
which allows the relevant algebraic manipulations and  analytic
 transformations.
The other important contribution from higher order processes
 corresponds to the QED rescattering processes
in which the scattering pion interacts with the electromagnetic field
by exchange of two virtual photons. The calculations of the
rescattering cross section lead to the analytic results presented
in~\cite{akhu2}.
 
These formulae were incorporated into the Fortran
code RCFORGV~\cite{akhu4}. The RCFORGV
program calculates more than 100 contributions from the
43 Feynman graphs, that are needed to describe the pion scattering
by nuclei: 3 diagrams in the Born approximation, one graph for the
structure
effects (polarizability), and 39 diagrams for the radiative corrections.
 The numerical calculations in the RCFORGV program are performed
by the Monte Carlo
 method. Double precision numbers are necessary for the calculations of
 the contribution of the double bremsstrahlung process (4).
 The program package RCFORGV
is set up on a IBM-3072 at Tel Aviv University.
 
The semi-analytical program RCFORGV calculates
 the radiative corrections of order {\oa} to the process (1)
 for the following cross sections:\\
 
 i) for $ d{\sigma} /  d{\omega} $
    in the $\omega$-bins : $\omega=E_1-E_2$;\\
 
 ii) for $ d^{2}{\sigma}/ dE_2dQ^2 $
     in the points of $(E_2,Q^2)$ .\\
 
 Here $Q^2= -(p_1-p_2)^2$, and $E_1~(E_2)$ are the initial (final) energies
 of the pions in the laboratory system. 
 These cross sections are needed to the analysis of planned experiments.
 
 In RCFORGV, it is easy to provide a variety of kinematic cuts and
criteria for event selection, and to take into account
the experimental geometry.
 The following kinematic constraints are included
 in the program:\\
 
 1) $ t=-k^2=-(p_1-p_2-k')^2 < \bar{t}   $
    for the 4-momentum transfer ;\\
 
 2) $ {\theta}_{\pi} < \bar{\theta}_{\pi} $
    for the pion scattering angle in the laboratory frame;\\
 
 3) $\omega>\bar{\omega} $
    for the photon detection threshold;\\
 
 4) ${\theta}_{\gamma\gamma} > \bar{\theta}_{\gamma\gamma}$
    for the angular resolution of the double bremsstrahlung.\\
 Here $\bar{t} $ is the maximum value of the squared momentum transfer,
 fixed by experimental resolution in $t$. Below $\bar{t}$, Coulomb 
 scattering dominates~\cite{GMOP}. Also, $\bar{\theta}_{\pi}$ is the
 maximum of the pion scattering angle.  
 The parameters $ \bar{\omega} $ and $\bar{\theta}_{\gamma\gamma} $
 are important for distinguishing double and single 
 bremsstrahlung.
 If  $\omega<\bar{\omega} $  or
     ${\theta}_{\gamma\gamma} < \bar{\theta}_{\gamma\gamma}$,
   then these two processes are indistinguishable~\cite{akhu2}.
 
Using RCFORGV, the  cross sections at the energies  40 and 600 GeV
for the radiative  scattering of $\pi^-$ mesons on carbon,
were calculated in the Born
approximation (the Born cross section), 
including polarizability contributions and
radiative corrections.
Numerical calculations were performed by the Monte Carlo method,
using 50 000 events for each incident energy. The value 40 GeV
was chosen to test  our new program installation
with results given previously in~\cite{akhu3}. One set
of calculations  was done for lead, to investigate the Z dependence.
The cross sections were calculated with kinematic
constraints given in Table 1.
\begin{table}
\begin{center}
\begin{tabular}{|c|c|c|c|c|} \hline
Energy        & \multicolumn{4}{c|}{ Parameters} \\
\hline
              & $\bar{t}$ & $\bar{\omega}$ & $\bar{\theta}_{\pi}$
              & $ \bar{\theta}_{\gamma\gamma}$\\
\hline
 $GeV$      & $GeV^2$ & $GeV$ & $rad$ & $rad$\\
\hline
40       & $2\times10^{-4}$ & 0.5 & $1.5\times10^{-2}$ & $2.0\times10^{-3}$ \\
\hline
600      & $4\times10^{-3}$ & 10.0 & $7.0\times10^{-3}$ & $2.0\times10^{-4}$ \\
\hline
\end{tabular}
\end{center}
\caption{{\it The kinematic constraints  used in the program
              RCFORGV.}}
\label{tabl1}
\end{table}

These constraints take into account the geometry and the event selection
criteria for experiments~\cite{Antip12} and~\cite{mur}.
Due to kinematics, we chose the parameters of $\bar{t}$ and
$\bar{\omega}$ larger, and angles
$\bar{\theta}_{\pi}$ and $\bar{\theta}_{\gamma\gamma}$ smaller, for
energy 600 GeV compared to 40 GeV.
 
\section
{The role of radiative correction
\label{RC role}}
 
 In this section we present the RCFORGV calculations.
 Following ~\cite{akhu2}, we define the radiative correction factor
$\delta(\omega)$ as the ratio of the radiative corrections cross section to
the Born cross section, as a function of the detected gamma energy $\omega$:
$$
\delta({\omega} )=
 \frac{(d {\sigma}/d {\omega })^{rc}}
      {(d {\sigma} /d {\omega })^{Born}}.
 \eqno(6)
$$

\begin{figure}[htb]
\centerline{\psfig{figure=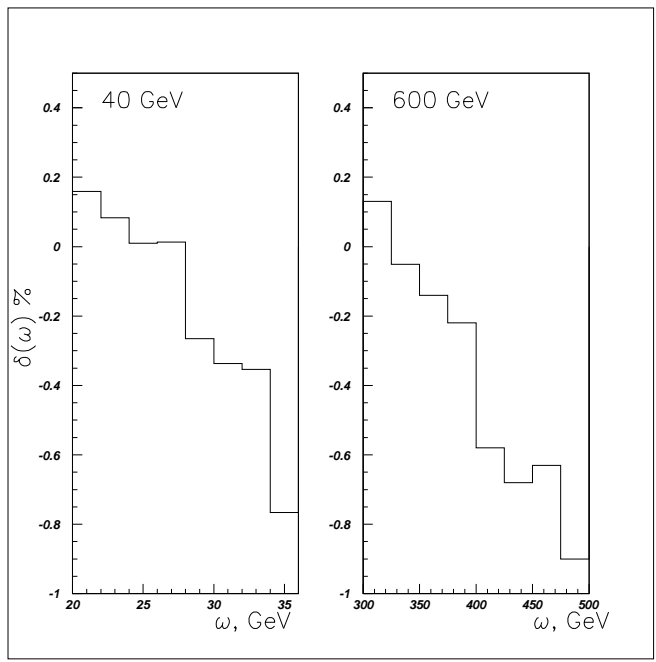,height=10cm}}
\caption{{\it Radiative correction $\delta(\omega)$ for radiative scattering
 of 40 and 600 GeV ${\pi}^{-}$-mesons on $C^{12}$ nuclei.}}
\label{fig2}
\end{figure}

 The results of the RC calculations for the radiative
  scattering of 40 GeV and 600 GeV ${\pi}^{-}$ mesons on carbon
 nuclei are presented in Fig.~\ref{fig2}.
 One sees that the absolute value of $\delta(\omega)$
 is less than 1\% over a large energy range.
 
\begin{figure}[htb]
\centerline{\psfig{figure=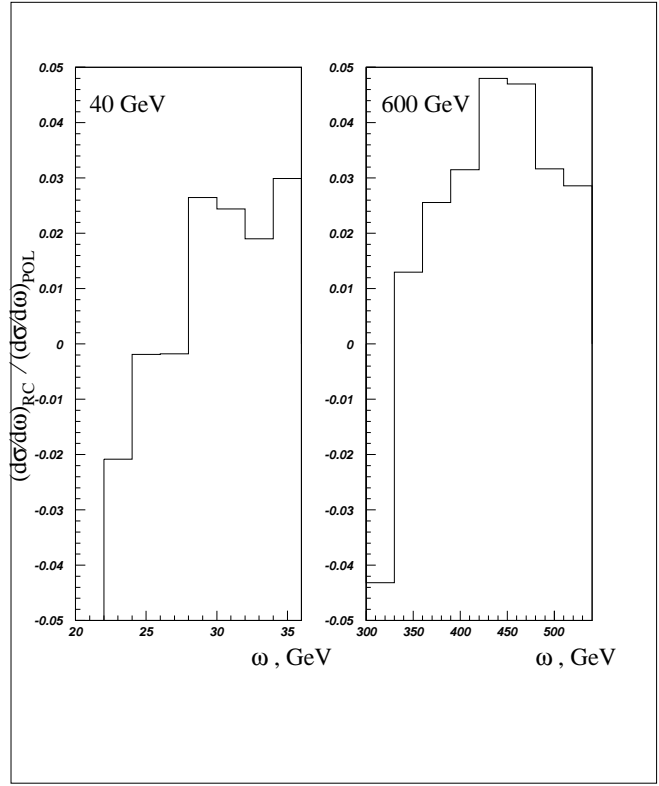,height=10cm}}
\caption{{\it Ratio of the radiative correction to the polarizability cross
section for radiative scattering
of 40 and 600 GeV ${\pi}^{-}$-mesons on $C^{12}$ nuclei.}}
\label{fig3}
\end{figure}

The polarizability cross section can be written
as the sum of two terms ~\cite{akhu3}:
$$
d {\sigma}^{pol} = - (\bar\alpha_{\pi} + \bar\beta_{\pi} ) d {\sigma^{e}} +
   \bar\beta_{\pi}d {\sigma^m} ,
 \eqno(7)
$$
where $d {\sigma^{e}}$ and $d {\sigma^m}$ give different combinations
of electric and magnetic polarizabilities.
We imposed the condition 
$\bar\alpha_{\pi} + \bar\beta_{\pi} = 0 $.
In our analysis, we therefore determine radiative corrections with respect to 
one polarizability value  $\bar\alpha_{\pi}=-\bar\beta_{\pi}$. 
To set the scale, we used the values:
$$
 \bar\alpha_{\pi}=-\bar\beta_{\pi}=5.
 \eqno(8)          
$$
The ratio of the radiative correction cross section to polarizability cross
section is presented in Fig.~\ref{fig3}. This ratio
changes in the interval $\sim( -5 ; 5)\%$ as function of gamma energy
$\omega$.
The formula (5) can be written:
$$
d {\sigma}^{tot} = d {\sigma}^{Born} +
                          \bar\beta_{\pi}d {\sigma}^{m} +
                        d {\sigma}^{rc}.
\eqno(9)
$$
To emphasize the fact that
radiative corrections can simulate a polarizability, we rewrite (9):
$$
d {\sigma}^{tot} = d {\sigma}^{Born} +
    ( \bar\beta_{\pi} + \beta_{\pi}^{rc})  d {\sigma}^{m}, 
 \eqno(10)
$$
where $ \beta_{\pi}^{rc} =  d {\sigma}^{rc}/d {\sigma}^{m}$.
 
In this analysis, we calculate also the double differential cross
section $ d^2{\sigma}/dE_2dQ^2$. We can transform this cross section to
the center mass of the ${\gamma}{\pi}$ system, as
 $d^2{\sigma}/ds_1dcos{\theta}^*$.
 Here, $s_1=(p_1+k)^2=(p_2+k')^2$
and $\theta^*$ are the $\gamma\pi$ energy squared, and the Compton
scattering angle.
At very small $ -k^2 ~\leq~\bar{t} $ and small
${\theta}_{\pi} ~\leq~\bar{{\theta}_{\pi}} $ we find:
$$
\frac{d^{2} {\sigma}}{ds_1 dcos{\theta^*}}=
  \frac{E_1 {(s_1-m_{\pi}^2)}^2 (1 - cos{\theta^*})}{4s_1^2}
  \frac{d^2 {\sigma}}{dE_2dQ^2}.
\eqno(11)
$$
 
\begin{figure}[htb]
\centerline{\psfig{figure=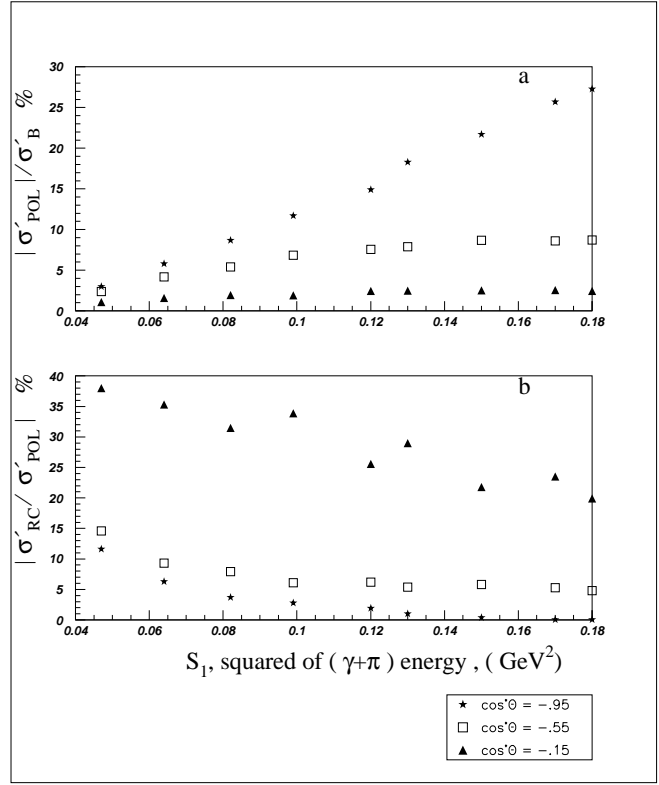,height=10cm}}
\caption{{\it Ratio of  $\sigma' \equiv d^2{\sigma}/dE_2dQ^2$ for:
a) the polarizability and the Born cross sections;
b) the radiative correction and the polarizability cross sections.}}
\label{fig4}
\end{figure}

The ratio of the polarizability to the Born double differential cross
section for
different scattering angles, as a function of $s_1$,  is
presented in Fig.~\ref{fig4}a.
In this figure ratios of $ d^2{\sigma}/dE_2dQ^2$ were taken, but the labeling
is according to the CM variables $s_1$ and $cos{\theta^*}$.
It is seen that the absolute value of the
polarizability contribution increases with energy and can reach a
value of $\sim30\%$ at backward angles. 
For a gamma energy
$\omega^*=(s_1 - m_{\pi}^2)/2\sqrt{s_1}\approx150 MeV$
 and $cos{\theta^*}\approx -0.5$,
the contribution
of magnetic polarizability to the total cross section of reaction (1) is
about $7\%$. At the point
$s_1=0.18 GeV^2$ ($\omega^*\approx 200 MeV$) and $cos{\theta^*}=-0.95$,
this contribution attains  magnitude of $\sim30\%$.
In the laboratory frame, this kinematical range corresponds
to the range of detected gamma energies  $\omega\sim(0.60-0.90)E_1$, where

$$
\omega =\frac{E_1}{2}(1-\frac{m_{\pi}^2}{s_1})(1-cos{\theta^*}).
\eqno(12)
$$
 
Here, we point out that data of~\cite{Antip12,Antip34} were taken
in this region of $\omega$ as well.
These authors showed that the preferred kinematical range to study
the magnetic polarizability in the laboratory frame is the
range of final gamma energy ${\omega}~\geq~0.75E_1$. In the
center of mass frame, this region is defined by the
inequalities:
$s_1\geq0.15 GeV^2$ and $cos{\theta^*}\leq-0.75$. One sees
from Fig.~\ref{fig4}a that for this kinematical range,
the relative contribution of magnetic
polarizability to the Born cross section is more than $15\%$.
 In Fig.~\ref{fig4}b, the ratio
of the RC to polarizability double differential
cross section is shown. We see that the relative contribution 
from radiative corrections to the polarizability cross section 
contribution is less than $5\%$ at backward
angles for large enough values of $s_1\geq0.13 GeV^2$.
 
We wish to determine
the magnitude of the simulated contribution to the polarizability,
due to the radiative
corrections. We can do this by calculating
the average value of the cross section ratio
$d {\sigma}^{rc}/d {\sigma}^{m}$ for the gamma energy range
$\omega\geq0.75E_1$. We have:
$$
{\beta_{\pi}^{rc}}=\frac{1}{n}\sum_{i=1}^n d {\sigma}^{rc}(\omega_i)/
d {\sigma}^{m}(\omega_i),
\eqno(13)
$$
where $n$ is the number of $\omega$ bins.
 
For the energies 40 and 600 GeV, we obtained
${\beta_{\pi}^{rc}}\approx -0.20 $.
The accuracy of our calculation may be estimated as:
$${\delta}({\beta_{\pi}^{rc}})\approx
\sqrt{n} {\epsilon}_{tot} {\beta_{\pi}^{rc}},
\eqno(14)
$$
where ${\epsilon}_{tot}=\sqrt{{\epsilon}_{rc}^2+{\epsilon}_{pol}^2}$.
Here, ${\epsilon}_{rc}$
and ${\epsilon}_{pol}$ are the relative uncertainties in the
 radiative correction
and polarizability cross section calculations. These uncertainties  
were calculated by the RCFORGV program 
for each of $\omega$ bins and we have found that over the  gamma
energy range  $\omega\geq0.75E_1$  all of ${\epsilon}_{tot}$ are 
approximately the same . This fact explains the appereance of 
$\sqrt{n}$ in (14). 
We have obtained  ${\epsilon}_{tot}\approx0.12$,
which gives
the calculation accuracy ${\delta}({\beta_{\pi}^{rc}})\approx0.05$.
 
\begin{figure}[htb]
\centerline{\psfig{figure=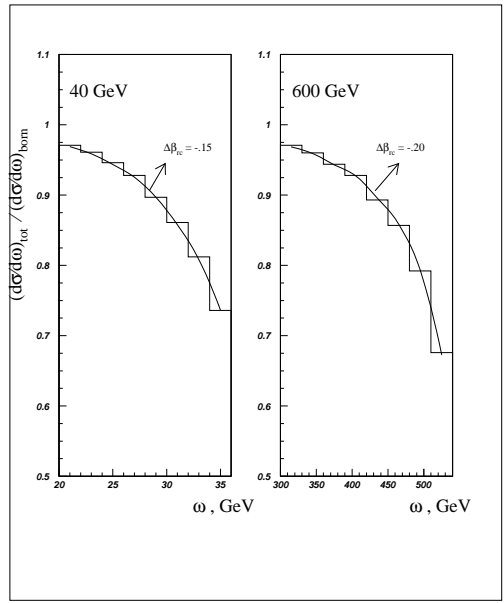,height=10cm}}
\caption{{\it The ratio of the total cross section to the Born cross section
as a function of detected gamma energy $\omega$.}}
\label{fig5}
\end{figure}

In Fig.~\ref{fig5}, we show the ratio of the total cross section
for reaction (1) to the Born cross section. This is given as
a function of gamma energy
$\omega$, for incident pion energies 40 and 600 GeV.
The histogram gives $d {\sigma}^{tot}$ 
calculated by (9) with  $\bar{\beta}_{\pi}=-5$, and the curve corresponds to 
$d {\sigma}^{tot}$  calculated by (10), with $\bar\beta_\pi=-5$ and  
${\beta_{\pi}^{rc}} = - 0.20$. It is clear that both calculations 
are in agreement. The value  of ${\beta_{\pi}^{rc}}$  also 
agrees with the results presented in \cite{akhu3}.
The same analysis was done assuming that 
$ \bar\alpha_{\pi}=\bar\beta_{\pi}= 0$, and we again find 
${\beta_{\pi}^{rc}}\approx -0.20$.  
This last result is important, showing that the 
radiative corrections simulate ${\beta_{\pi}^{rc}}\approx -0.20$, 
independent of the $\bar\beta_{\pi}$ value.

One calculation with $\bar\beta_\pi=-5$ was carried out to
study radiative corrections to pion radiative scattering on lead. 
The Born and polarizability
cross sections show the $Z^2$ dependence expected theoretically.
The ratio of radiative correction  to Born 
and to polarizability
cross sections was practically unchanged compared to carbon.
This is because ${\delta}({\omega})$ is practically independent of Z.
Only the contribution from rescattering has a ~Z dependence, and this is 
a very small contribution to the total RC.

 We now summarize the  main results of our RC study.  
The radiative corrections simulate a polarizability effect at
the level of ${\beta_{\pi}^{rc}} \sim -0.20$.
This result is for pion incident energies from 40 to 600 GeV, and for 
nuclei from carbon to lead. We find for 
${\beta_{\pi}^{rc}}\approx - 0.20 \pm0.05$ at 40 and 600 GeV,
for $\bar\beta_{\pi}$ from the interval $(-7; -2)$, 
the relative contribution to the polarizability from 
radiative corrections equals ${\beta_{\pi}^{rc}} /\bar\beta_{\pi}
\approx (3-10)\%$.
 Therefore, we give a simple lowest order method of obtaining the 
polarizability from experimental data. We write the 
cross section as a sum of Born and polarizability cross sections:
$$
d {\sigma}^{exp} = d {\sigma}^{Born} +
      \bar\beta_{\pi}^{exp}    d {\sigma}^{m}.
 \eqno(15)
$$
We now find the best fit $\bar\beta_{\pi}^{exp}$ for $d {\sigma}^{exp}$ 
using RCFORGV, or any other convenient fit program.
This experimental $\bar\beta_\pi^{exp}$ must be corrected by
${\beta_{\pi}^{rc}}$.
We have by comparison to (10), the relationship:
$\bar\beta_{\pi}^{exp} = \bar\beta_{\pi} - {\beta}_{\pi}^{rc}$.  
Thus a value $\bar\beta_{\pi}^{exp}=-5.20$ would 
correspond a polarizability $\bar\beta_{\pi}=-5$. 
A more exact determination would require 
analysis of the data with RCFORGV. 

\section
{Different Born cross section calculations
\label{epaww}}
 
In this section, we briefly describe two methods for the
Born cross section calculation. The first is an exact
calculation performed with the help of the RCFORGV program,
and the second is the equivalent photon approximation (EPA)
(Weizs\"acker-Williams's method) ~\cite{WW,Akhieser}.
 
The exact calculation method consists of an accurate counting of the
contributions from the all QED diagrams in the lowest order of $\alpha$.
This corresponds to the scattering of a spinless pointlike particle by a
Coulomb center with hard photon emission.
The exact calculation takes  into account the
magnitude of the photon virtuality $k^2$.
The exact expression of the Born cross
section of the process (1) has the following form \cite{akhu2}:
$$
\frac { d^2{\sigma}^{Born}}{dE_2dQ^2} = \frac{Z^2{\alpha}^3}
                        {{\vec{p_1}}^2}
                          \int_{t_{min}}^{t_{max}}\frac{dt}{t^2}S^0(t).
\eqno(16)
$$

The explicit form of the function $S^0(t)$ and the kinematical limits
$t_{min,max}$ are given in~\cite{akhu1,akhu2}.
Here we emphasize that the exact Born cross section calculation
includes the contribution
of the logarithmic term $\sim ln(t_{max}/t_{min}) $
(leading logarithmic approximation),
and the additional contribution of the next term which is proportional
to $\sim (1/t_{min}-1/t_{max})$.
 
The EPA method is based on the assumption that the photon virtuality
is very small,  $-k^2 ~\leq~ \bar{t} << m_{\pi}^2 $.
Therefore, one ignores the small  deviation from the real photon,
and one takes only the leading logarithmic term.
In this case, the cross section of the reaction (1) can be
expressed via the cross section for the $\gamma\pi$ scattering (2):
$$
   d \sigma _{{\pi}Z\rightarrow{\pi}{\gamma}Z}  =
   n(s_1)d s_1 d \sigma_{\gamma\pi}(s_1).
\eqno(17)
$$
Here, $ d \sigma_{\gamma\pi}(s_1)$ is  the cross section of the
Compton scattering of the real photon, $ n(s_1)$ is
the equivalent photon density .
 
In the EPA, the cross section for the reaction (1) in the center of mass frame
can be written:
$$
\frac {d^2{\sigma}^{WW}}{ds_1dcos{\theta^*}}
             =\frac{\alpha}{\pi}\frac{Z^2}{s_1-m_{\pi}^2}
   \left[ \ln \frac{t_{max}}{t_{min}} +\frac{t_{min}}{t_{max}}
                - 1 \right]
        \frac{ d{\sigma}_{\gamma\pi}}{dcos{\theta^*}} (s_1),
\eqno(18)
$$
\noindent
where $ d{\sigma}_{\gamma\pi}(s_1)/dcos{\theta^*}$
is the unpolarized differential cross section of the reaction (2).
 
The Born cross section of the $\gamma\pi$ Compton scattering
has the form~\cite{babu2} :
$$
    \frac{d \sigma^{Born}_{\gamma\pi}}{dcos{\theta^*}}(s_1)
    = \frac{\pi\alpha^2}{s_1}
       \left[{1 + [-1 +
       \frac{2m_{\pi}^2 t_1}{(s_1-m_{\pi}^2)(u_1-m_{\pi}^2)} ]^2} \right],
\eqno(19)
$$
 where $t_1=(p_1-p_2)^2, u_1=(p_1-k')^2$.
\begin{figure}[htb]
\centerline{\psfig{figure=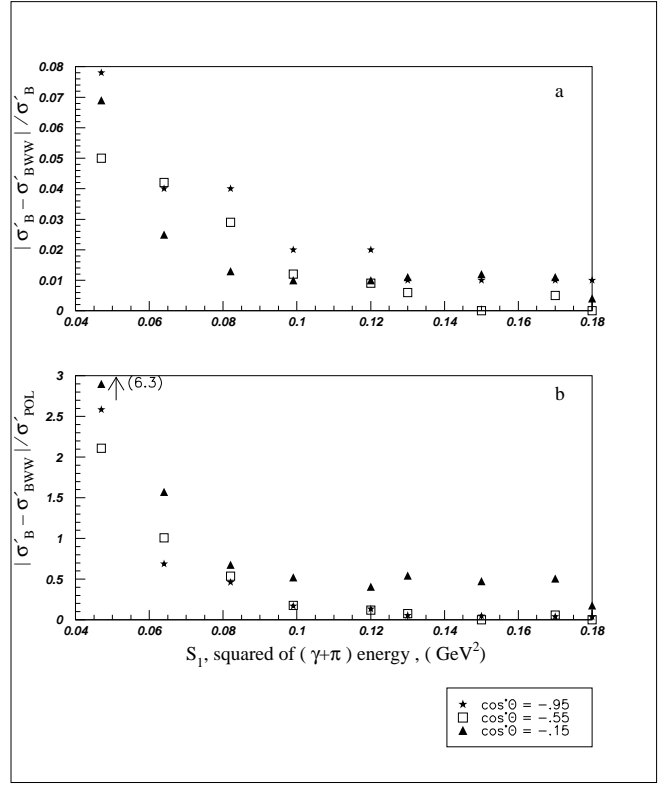,height=10cm}}
\caption{ { \it Comparison of two Born cross section calculations:
exact calculation  and EPA method
($\sigma' \equiv d^2{\sigma}/dE_2dQ^2$).}}
\label{fig6}
\end{figure}

A comparison between the two Born cross section
calculations was carried out 
in order to estimate their applicability.
In Fig.~\ref{fig6}a the difference between the two double differential
cross sections  $ d^2{\sigma}/dE_2dQ^2$, is presented.
The superscript ${Born}$ refers  to the exact Born cross
section performed by RCFORGV program, and  ${WW}$ to the Born
cross section calculated by the EPA method.
As in Fig.~\ref{fig4}, ratios of $ d^2{\sigma}/dE_2dQ^2$ were taken, but
the labeling is according to the CM variables $s_1$ and $cos\theta^*$.
The cross section difference
is given with respect to the exact Born cross section.
 
From  Fig.~\ref{fig6}a, we can see that
the EPA method agrees with the exact Born calculation with
a precision better then 3 \% in the kinematical
region $s_1\geq0.13  GeV^2$ and at backward scattering angles
$cos\theta^*\leq-0.5$.
From Fig.~\ref{fig6}b, the cross sections difference
divided by the polarizability cross section is displayed.
In the kinematical range limited by the
inequalities  $s_1\geq0.13  GeV^2$ and $cos\theta^*\leq-0.5$,
both Born cross section calculation methods are in
agreement and can be applied equally well.
This kinematical region was also preferred
for the study of the magnetic polarizability (see section 3).
 
The validity of the EPA method was discussed in recent papers
(see for example~\cite{Budnev}-\cite{drees}) 
shown that the EPA for the other processes can lead to large errors,
if used outside of the region of applicability and must be used 
judiciously or improved in order to get correct results.
 
\section
{Conclusions
\label{concl}}
We used the semi-analytical program RCFORGV to evaluate the
contribution of the radiative corrections to the total cross section
of the pion scattering by a Coulomb center with hard photon
emission. We showed that the radiative corrections
can simulate the polarizability effect and the average 
effect $\beta_{\pi}^{rc}=(-0.20 \pm0.05)\times10^{-43} cm^3$ was obtained.
We showed that the preferred kinematical region to investigate
the magnetic polarizability cross section in the laboratory frame
is the range of final gamma energy $\omega\geq0.75E_1$ .
The corresponding range in the center of mass frame is defined by the
following inequalities : $s_1\geq0.15  GeV^2$ and
$cos\theta^*\leq-0.75$, where $s_1$ and $\theta^*$ are the energy squared and
the polar scattering angle.
The cross sections of the reaction (1) computed in the Born approximation
using the exact calculation method and the equivalent photon approximation
are in the agreement for hard emitted photons $\omega\geq0.75E_1$ .
In the range of the photons $\omega\leq0.6E_1$, we must use
the exact calculation method.
The suggested method allows to take into account the radiative correction
effects with high accuracy.

\section*{Acknowledgements}
 
We would like to thank D.~Bardin, E.~Gurvich, G.~Mitsel'makher,
A.~Ol'shevsky and L.~Frankfurt for numerous helpful discussions.
A.~A. thanks the Tel Aviv University group for the
opportunity to visit and work in Tel Aviv on this project,
and Professor Abdus Salam, the International Atomic
Energy Agency and UNESCO for hospitality at the International Centre for
Theoretical Physics, Trieste.
 
This work was supported by the
Yuval Ne'eman chair of Theoretical Physics at Tel Aviv University,
the Ministry of Absorption, the Israel Ministry of Science,
the Wolfson Foundation and
the U.S.-Israel Binational Science Foundation (B.S.F.), Jerusalem, Israel.

%


\newpage
\begin{thebibliography}{99}
\bibitem {kle} A.~ Klein, {\it Phys. Rev.} {\bf  99} (1955) 998.
\bibitem {balda}  A.~ M.~Baldin, {\it Nucl. Phys.} {\bf 18} (1960) 310.
\bibitem {petr1}
 V.~A.~ Petrun'kin, {\it Sov. J. Part. Nucl.} {\bf 12} (1981) 278.
\bibitem {lvov3} A.~I.~L'vov, {\it Sov. J. Nucl. Phys.} {\bf 42} (1985) 583.
\bibitem {hols1} B.~R.~Holstein, {\it Comments Nucl. Part. Phys.} {\bf 19}
(1990) 239.
\bibitem {dono} J.~F.~Donoghue and B.~R.~Holstein, {\it Phys. Rev.} {\bf
40D} (1989) 2378.
\bibitem{GMOP}
 A.~S.~Galperin et al., {\it Sov. J. Nucl. Phys.} {\bf 32} (1980) 545.
\bibitem{Antip12}
 Yu.~M. Antipov et al., {\it JETP Letters} {\bf 35} (1982) 375;\\
 Yu. M. Antipov et al., {\it Z. Physik} {\bf C24} (1984) 39.
\bibitem{Antip34}
 Yu. M. Antipov et al., {\it Phys. Letters} {\bf B121} (1983) 445;\\
 Yu. M. Antipov et al., {\it Z. Physik} {\bf C26} (1985) 495.
\bibitem{mur} M.~A.~Moinester, in: W.~Van Oers (eds.), Proceedings of the
Conference on the Intersections between Particle and Nuclear Physics,
Tucson, Arizona, 1991 (AIP Conference Proceedings 243, 1992) p.553.
\bibitem{akhu1}
A. A. Akhundov and D. Yu. Bardin, JINR Dubna preprint P2-82-650 (1982).
\bibitem{akhu2}
 A. A. Akhundov, D. Yu. Bardin and G. V. Mitsel'makher,
   {\it Sov. J. Nucl. Phys.} {\bf 37} (1983) 217.
\bibitem{akhu3}
 A. A. Akhundov, D. Yu. Bardin, G. V. Mitsel'makher and A. G. Ol'shevsky,
  {\it Sov. J. Nucl. Phys.}  {\bf 42} (1984) 426.
\bibitem{akhu4}
A. A. Akhundov and D. Yu. Bardin, Fortran program RCFORGV
 and JINR Dubna preprint P2-82-650 (1982).
\bibitem{SCHOONSCHIP}
M.~Veltman, SCHOONSCHIP - {\it A Program for Symbol Handling};\\
H.~Strubbe, {\it Comp. Phys. Comm.} {\bf 8} (1974) 1.
\bibitem{WW} K.~F.~von~Weizs\"acker, {\it Z. Physik} {\bf 88} (1934) 612;\\
E.~J.~Williams, {\it Phys. Rev.} {\bf 45} (1934) 729.
\bibitem{Akhieser}
 A.~I.~Akhieser, V.~M.~Berestetski,
  {\it Quantum electrodynamics} (Nauka, Moscow, 1981).
\bibitem {babu2} D. Babusci, S. Bellucci, G. Giordano, G. Matone,
A. M. Sandorfi and M. A. Moinester, {\it Phys. Letters} {\bf B277} (1992)
158.
\bibitem{Budnev}
 V.~M. ~Budnev et al., {\it Phys. Rep.} {\bf C15} (1975) 181.
\bibitem{frix} S.~Frixione et al, CERN preprint CERN-TH.7032/93 (1993).
\bibitem{drees} M. Drees and R. M. Godbole, Wisconsin University
 preprint MAD-PH-819, BU-TH-94-02 (1994).
 
\end{thebibliography}
\end{document}